\documentclass[sigconf,nonacm]{acmart}

\setcopyright{none}
\renewcommand\footnotetextcopyrightpermission[1]{}
\usepackage{enumitem}
\usepackage{svg}
\usepackage{url}

\AtBeginDocument{%
  }

\title{GALA: Multimodal Graph Alignment for Bug Localization in Automated Program Repair}

\author{Zhuoyao Liu}
\authornote{These authors contributed equally to this work.}
\affiliation{
  \institution{Sichuan University}
  \city{Chengdu}
  \country{China}
}
\email{liuzhuoyao@stu.scu.edu.cn} 

\author{Zhengran Zeng}
\authornotemark[1]
\affiliation{
  \institution{Peking University}
  \city{Beijing}
  \country{China}
}
\email{zhengranzeng@stu.pku.edu.cn}

\author{Shudong Huang}
\affiliation{
  \institution{Sichuan University}
  \city{Chengdu}
  \country{China}
}
\email{huangsd@scu.edu.cn} 

\author{Yang Liu}
\affiliation{
  \institution{Sichuan University}
  \city{Chengdu}
  \country{China}
}
\email{liuyyy111@gmail.com} 

\author{Shikun Zhang}
\affiliation{
  \institution{Peking University}
  \city{Beijing}
  \country{China}
}
\email{zhangsk@pku.edu.cn}

\author{Wei Ye}
\authornote{Corresponding author.}
\affiliation{
  \institution{Peking University}
  \city{Beijing}
  \country{China}
}
\email{wye@pku.edu.cn}

\begin{document}

\begin{abstract}
Large Language Model (LLM)-based Automated Program Repair (APR) has shown strong potential on textual benchmarks, yet struggles in multimodal scenarios where bugs are reported with GUI screenshots. Existing methods typically convert images into plain text, which discards critical spatial relationships and causes a severe disconnect between visual observations and code components. Consequently, localization degrades into imprecise keyword matching. To bridge this gap, we propose \textbf{GALA} (Graph Alignment for Localization in APR), a framework that shifts multimodal APR from implicit semantic guessing to explicit structural reasoning. GALA operates in four stages: first, it constructs an \emph{Image UI Graph} to capture visual elements and their structural relationships. Second, it performs \emph{file-level alignment} by cross-referencing this UI graph with repository-level structures (e.g., file references) to locate candidate files. Third, it conducts \emph{function-level alignment} by reasoning over fine-grained code dependencies (e.g., call graphs) to precisely ground visual elements to corresponding code components. Finally, conditioned on the aligned files and functions, it performs patch generation within the grounded code context. By systematically enforcing both semantic and relational consistency across modalities, GALA establishes a highly accurate visual-to-code mapping. Evaluations on the SWE-bench Multimodal benchmark demonstrate that GALA achieves state-of-the-art performance, highlighting the effectiveness of hierarchical structural alignment.
\end{abstract}

\maketitle

\section{Introduction}

\begin{figure}[t]
\centering
\includegraphics[width=1\columnwidth]{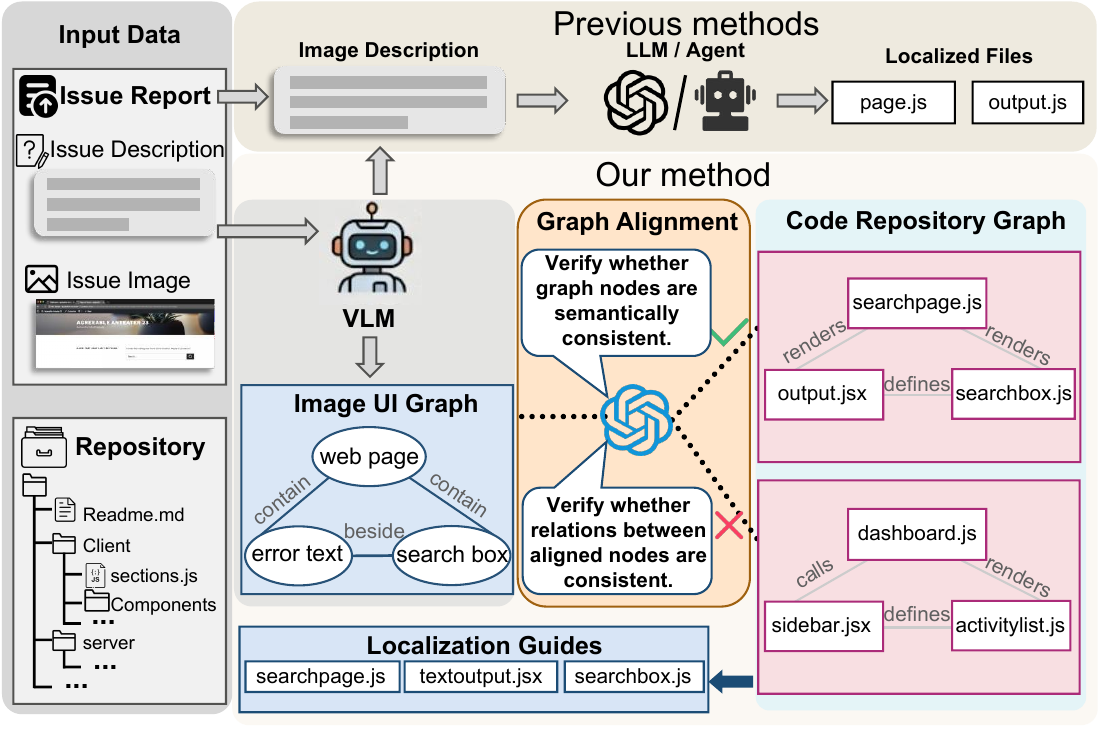} 
\caption{Comparison between previous works on Multimodal APR and our proposed method.}
\label{fig1}
\end{figure}

Automated Program Repair (APR) \cite{le2019automated,zhang2024systematic,zhang2023survey} aims to automatically identify and fix software defects, reducing human effort and improving software reliability. With the rapid advancement of large language models (LLMs), recent APR systems have achieved strong performance on real-world benchmarks\cite{tan2017codeflaws,ouyang2024benchmarking,lin2017quixbugs} such as SWE-bench\cite{jimenez2023swe}. However, these approaches are predominantly designed for unimodal settings, where both problem understanding and localization rely solely on textual inputs, including issue descriptions and source code.

In modern software development, especially for front-end systems, bug reports often include visual artifacts such as graphical user interface (GUI) screenshots. These visual signals provide critical information about layout, rendering, and interaction issues that cannot be fully captured by text alone. To address this, recent multimodal APR methods \cite{huang2025seeing,tang2026svrepair} attempt to incorporate visual inputs. As illustrated in Figure~\ref{fig1} (top), their typical workflow leverages Vision Large Language Models (vLLMs) to translate bug screenshots into natural language descriptions. These textual descriptions are then injected as supplementary context into traditional, single-modal fault localization and repair pipelines. However, we argue that this paradigm suffers from two fundamental limitations:
\begin{itemize}[leftmargin=*, nosep]
    \item \textbf{Loss of visual structural relationships:} Translating images to text naturally discards the complex spatial and structural relationships among UI elements. For example, consider a UI issue where a search box overlaps with a text label. A caption-based approach may correctly recognize the keywords “search box” and “text label,” but fails to capture the critical "overlap" interaction that defines the layout bug.
    \item \textbf{Inaccurate localization from visual-code disconnect:} Existing methods rely on implicit semantic keyword matching rather than structurally linking visual observations to the codebase. Consequently, given the previous overlapping issue, the model might retrieve unrelated files with similar semantics (e.g., text rendering utilities) instead of targeting the actual layout components responsible for the bug across the file and function hierarchies.
\end{itemize}

To address these limitations, we propose \textbf{GALA} (\textbf{G}raph \textbf{A}lignment for \textbf{L}ocalization in \textbf{A}PR), a framework that formulates multimodal localization as a structured cross-modal alignment problem. As shown in Figure~\ref{fig1} (bottom), our method consists of three key stages designed to systematically bridge the modalities.

First, to explicitly preserve visual structures (addressing Limitation 1), we construct an \emph{Image UI Graph}. By leveraging a Vision-Language Model (vLM) guided by the issue description, we extract key UI elements as graph nodes and their spatial or interactive relationships as edges, effectively capturing the topological context of the bug.
Second, to overcome the visual-code disconnect (addressing Limitation 2), we perform \emph{file-level alignment}. Instead of relying on isolated text matching, we model the repository context by extracting file paths and their inter-file reference relations. By feeding the textualized UI graph alongside this repository-level structural information into a Large Language Model (LLM), the model is able to cross-reference visual interactions with architectural dependencies to select candidate files.
Third, we conduct \emph{function-level alignment} within the selected files. We construct fine-grained code graphs containing function signatures and function call graphs. By jointly reasoning over the UI graph and these function-level structures, the LLM precisely grounds the visual elements to specific executable components.

Ultimately, by explicitly modeling structured representations across modalities and granularities, GALA transforms multimodal APR from implicit semantic guessing to explicit structural reasoning. This produces a highly accurate set of edit targets for downstream patch generation. Evaluated on the SWE-bench Multimodal benchmark, GALA achieves state-of-the-art performance, demonstrating the effectiveness of structure-aware alignment.

Our contributions are summarized as follows:
\begin{itemize}[leftmargin=*, nosep]
    \item We reformulate multimodal bug localization as a hierarchical cross-modal alignment problem across visual and code structures.
    \item We propose an image UI graph and multi-level code graphs to explicitly model both visual semantics and code dependencies.
    \item We design a hierarchical graph alignment mechanism that bridges visual elements to code components from file-level to function-level with both semantic and relational consistency.
    \item We demonstrate that our approach achieves superior performance on SWE-bench Multimodal, validating the effectiveness of structure-aware alignment for multimodal APR.
\end{itemize}

\section{Related Work}

\subsection{LLM for Code Localization}

Recent advances in Large Language Models (LLMs) have substantially improved bug localization by enabling deeper semantic understanding of both natural language issue descriptions and source code. Existing approaches can be broadly categorized into training-based models and prompt-based reasoning methods.

Training-based approaches learn to align bug reports with relevant code artifacts through supervised learning. For example, DNNLOC \cite{lam2017bug} integrates multiple handcrafted and learned features, while FBL-BERT \cite{ciborowska2022fast} adopts a ColBERT-style late interaction mechanism to capture fine-grained token-level relevance. BLAZE \cite{chakraborty2025blaze} further enhances retrieval via contrastive learning and dynamic chunking. Other neural approaches also explore learned representations for bug localization, highlighting the importance of fine-grained semantic alignment between bug reports and code \cite{ali2023automated,huo2017enhancing}. However, these methods depend on task-specific training and frequent retraining, limiting scalability in evolving codebases.

Prompt-based approaches instead leverage the reasoning capability of LLMs without additional training, framing localization as a combination of retrieval and structured reasoning. Some methods directly rank relevant files or code snippets based on textual similarity \cite{reddy2025swerank}. Building on this, recent work introduces iterative reasoning that progressively refines candidate sets through multi-step exploration of repository contexts \cite{jiang2025issue,xu2026learning}. Agent-based frameworks further extend this paradigm by enabling tool use, allowing LLMs to perform code search, file inspection, and hypothesis verification \cite{chen2025locagent,batole2025llm,samir2026improved,li2025swe}. To better capture repository-scale dependencies, several methods incorporate long-context modeling or repository-level memory \cite{wang2025extracting,wang2025improving}, while others introduce structural representations such as code graphs to guide search and constrain the solution space \cite{liu2025graphlocator}. However, these approaches remain largely unimodal and fail to capture structured visual semantics, limiting their effectiveness for multimodal software issues.

\begin{figure*}[t]
    \centering
    \includegraphics[width=\textwidth]{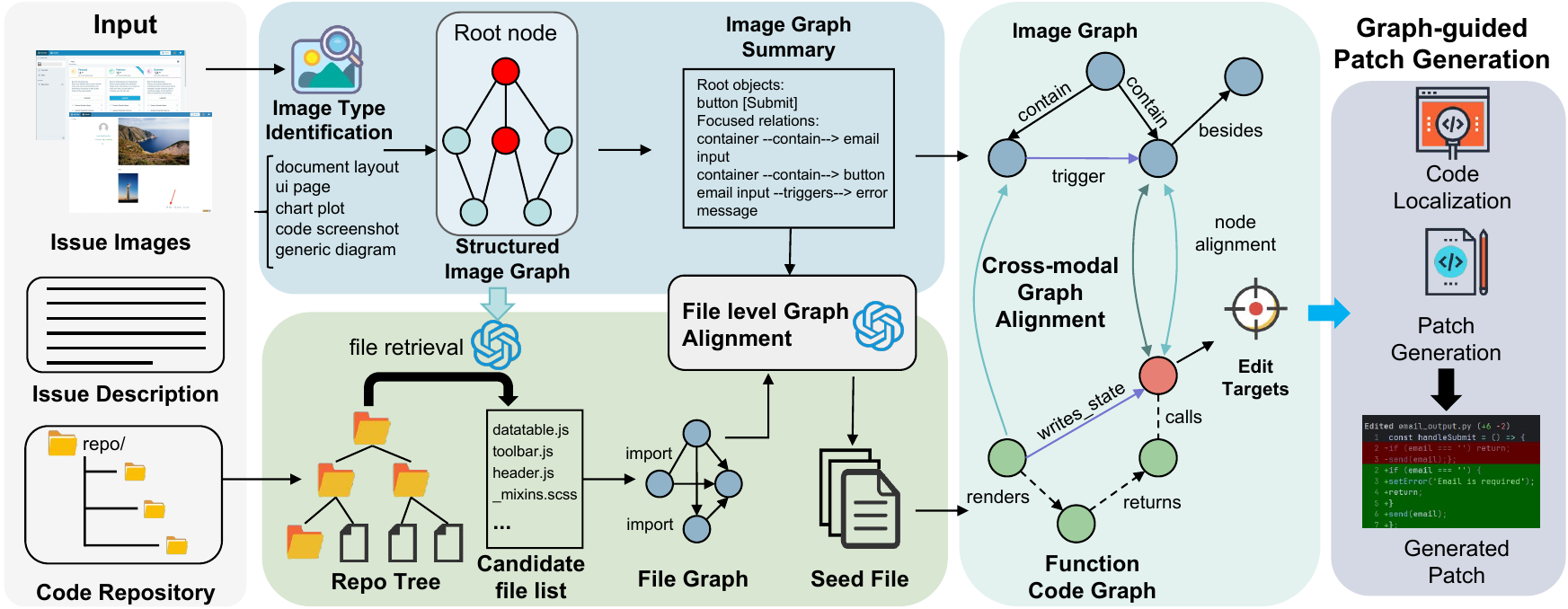}
    \caption{Architecture of GALA. The overall workflow consists of four key stages: (1) the Image Graph Construction module, which converts issue images into a problem-centric structured graph via type-aware parsing and rooted expansion; (2) the File-level Alignment module, which grounds visual semantics to the repository and identifies a compact set of seed files through structure-aware refinement; (3) the Function-level Alignment module, which performs cross-modal graph alignment to establish node- and relation-consistent correspondences between visual elements and code components, producing precise edit targets; and (4) the Graph-guided Patch Generation module, which leverages the aligned graph structure to constrain localization and generate minimal yet complete patches.}
    \label{method}
\end{figure*}

\subsection{LLM/MLLM for APR}

LLMs have also been widely explored for automated program repair (APR). Early approaches focus on function-level repair through fine-tuning \cite{jiang2023impact,xia2023automated,wu2023effective, wang2023rap,huang2025template,xia2023plastic} or prompting \cite{xia2022less,fan2023automated,zhao2024enhancing,bouzenia2025repairagent,lee2025unidebugger,yin2024thinkrepair,zhang2023gamma,yang2024cref,peng2024domain}. Recent works shift toward repository-level issue resolution, where agent-based frameworks enable LLMs to interact with execution environments and handle complex tasks \cite{zhang2024autocoderover,ma2025alibaba,antoniades2024swe,xia2025demystifying}. For example, SWE-agent \cite{yang2024swe} builds on SWE-Bench \cite{jimenez2023swe} and adopts an interactive agent paradigm for end-to-end issue resolution. Despite strong performance, these approaches are designed for unimodal settings and lack mechanisms to incorporate visual evidence.

A growing line of work extends APR to multimodal settings by incorporating visual inputs. Several emerging benchmarks have also explored multimodal software reasoning scenarios\cite{li2024mmcode,wang2025code}, highlighting the increasing importance of visual understanding in APR. Early frameworks such as Agentless \cite{xia2024agentless} and SWE-agent \cite{yang2024swe} have been adapted to SWE-bench Multimodal, demonstrating competitive performance. OpenHands-Versa \cite{soni2025coding} further integrates visual observations into the repair loop for grounded reasoning. GUIRepair \cite{huang2025seeing} leverages vision-language models to translate visual artifacts into executable reproduction scripts for localization and patch generation. SVRepair \cite{tang2026svrepair} advances this direction by incorporating structured visual cues to improve performance. However, existing methods still rely on implicit or weakly structured visual representations, failing to explicitly model fine-grained UI elements and their relational dependencies. This leads to the loss of critical visual context and limits precise cross-modal reasoning. In contrast, our approach constructs a structured intermediate representation that explicitly models UI elements and their structural and relational dependencies, enabling more accurate cross-modal alignment between visual evidence and code semantics, and improving both localization and repair.

\section{Method}

\subsection{Overview}

We propose GALA, a graph-based framework for multimodal bug localization and repair that bridges visual symptoms and code semantics through hierarchical graph alignment. As illustrated in Figure~\ref{method}, given an issue description, its corresponding issue images, and the code repository, GALA follows a four-stage pipeline. First, we construct a problem-centric image graph to capture structured visual elements and their relationships. Second, we perform file-level alignment to identify a compact set of candidate files by grounding visual semantics into the repository structure and refining them into seed files. Third, we conduct function-level alignment within the selected files, where visual elements are grounded to fine-grained code components through cross-modal graph alignment, producing precise edit targets. Finally, we perform graph-guided patch generation, where the repair process is constrained by the aligned graph structure, enabling targeted and reliable code modification. Each stage of this pipeline is instantiated through stage-wise simplified prompts, as illustrated in Figure~\ref{method}.

\subsection{Image Graph Construction}

\begin{figure}[t]
    \centering
    \includegraphics[width=1\linewidth]{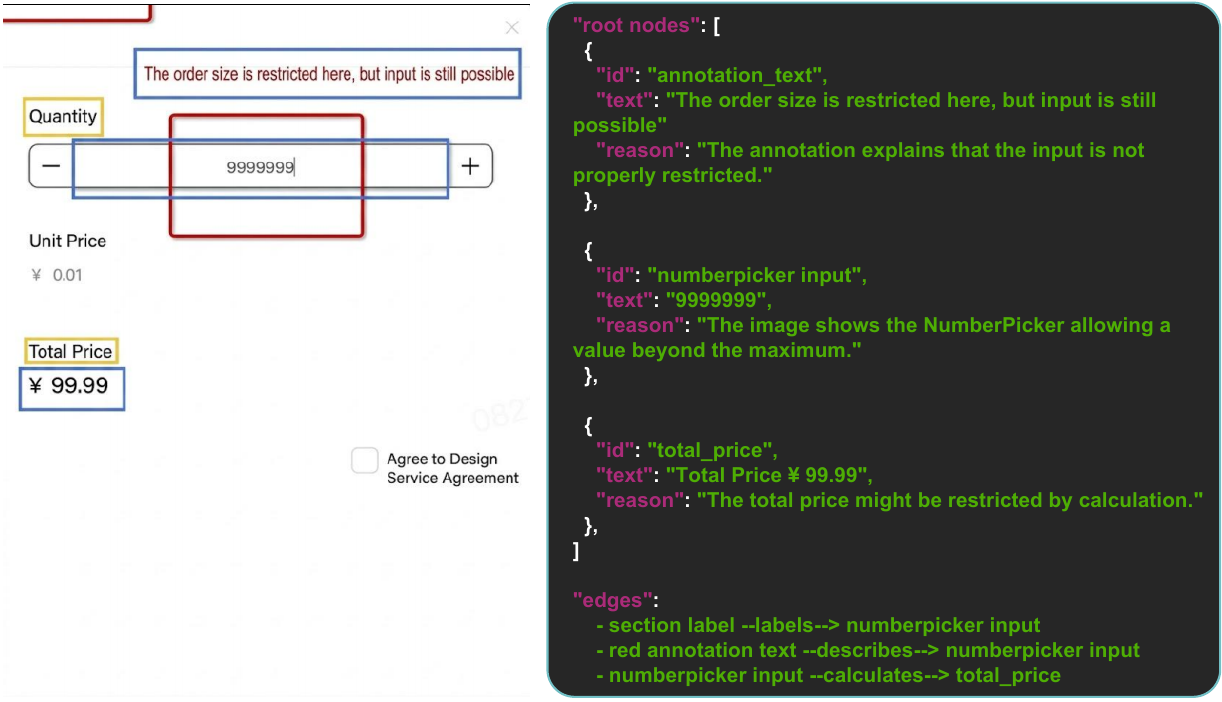}
    \caption{Example of the image graph constructed by GALA from alibaba-fusion\_next-4021(left: input image; right: image graph)}
    \label{image_graph}
\end{figure}

To enable structured visual reasoning, we construct an Image Graph that represents the input image as a problem-centric structured graph, as illustrated in Fig.~\ref{image_graph}. Formally, the image graph is defined as \(G_v=(V_v,E_v)\), where nodes \(V_v\) denote issue-relevant visual entities and edges \(E_v\) denote relations that are necessary for explaining the reported problem. Instead of directly extracting a global scene graph, our approach performs a type-aware and rooted graph construction process that retains only issue-relevant elements and their relations. This design is motivated by the observation that global visual parsing often introduces a large amount of irrelevant context, while bug understanding typically depends on a small set of semantically critical elements and their interactions.

\textbf{Image Type Identification.}
Given an input image and the corresponding issue description, we first identify the high-level visual type of the image using a vision-language model. Specifically, the model predicts one of five predefined categories—\textit{ui page}, \textit{chart plot}, \textit{code screenshot}, \textit{document layout}, or \textit{generic diagram}. Outputs a discrete image type label to guide subsequent graph construction. This step establishes a type-specific structural prior. Different image types exhibit distinct structural patterns and salient elements (e.g., UI components in web pages versus data entities in charts), making a unified extraction strategy prone to introducing irrelevant structures. By conditioning on the predicted type, the model adapts its perception strategy to focus on task-relevant visual semantics, while the predicted type also constrains the candidate node categories and admissible relation patterns for graph construction.

\textbf{Root Object Selection.}
We use a vision-language model (VLM) to identify a small set of \textit{root objects} that serve as anchors of the graph, conditioned on the issue image, issue description, and the predicted image type from the previous stage. Conceptually, root object identification, supporting node expansion, and relation construction are defined as three constrained substeps, although they can be instantiated within a single structured VLM generation for efficiency. Root objects correspond to elements that are directly related to the issue, including: (i) objects affected by the bug, (ii) objects visually involved in the reported abnormality, and (iii) objects explicitly referenced in the issue description. As illustrated in Fig.~\ref{image_graph}, these root objects are highlighted by blue boxes and capture the key elements involved in the bug. We prioritize objects supported by both visual evidence and textual description, while also allowing visually evident abnormal objects when the textual description is underspecified. To improve reliability, each selected root object must be explicitly grounded in observable visual content and justified as issue-relevant; candidates lacking clear support are discarded. Each root object is associated with a justification explaining its relevance to the issue. This step establishes the semantic core of the graph and enforces a rooted structure, where subsequent nodes are introduced only to explain these anchors, preventing the graph from expanding into irrelevant visual regions.

\textbf{Supporting Node Expansion.}
Conditioned on the identified root objects, the model further introduces additional nodes that are necessary to explain the issue within the same generation process. These \textit{supporting nodes} provide minimal contextual information required for understanding the problem, including local context (e.g., neighboring UI components), structural context (e.g., containers or boundaries), and objects that participate in issue-relevant relations. As shown in Fig.~\ref{image_graph}, these supporting nodes are highlighted by yellow boxes and provide the minimal context needed to interpret the relationships among root objects. Rather than aiming for exhaustive scene coverage, node expansion is restricted to one-hop issue-relevant context that is indispensable for explaining the retained root objects and relations, ensuring that the resulting graph remains a localized, problem-centric substructure instead of a global scene representation. This design significantly reduces graph size and suppresses irrelevant visual context, which is critical for stabilizing downstream reasoning. Each node is accompanied by a textual rationale describing its role in the explanation.

\textbf{Relation Construction.}
Within the same generation process, the vision-language model (VLM) further infers directed edges between the identified nodes to capture issue-relevant structural and functional relations. Each edge is defined as \(e_{ij} = (v_i, r, v_j)\), where \(r\) denotes a relation label generated by the model according to the local visual structure and issue context. Edges are added only when the inferred relations are visually grounded and contribute to understanding the reported issue; uncertain or weakly supported relations are omitted. Rather than constructing a dense graph over all detected entities, relation generation is restricted to the smallest connected issue-relevant substructure centered on the most relevant UI components and their associated stateful elements. Consequently, the resulting graph preserves sparse but informative dependencies that support downstream code alignment. Each relation is further associated with a justification explaining both its visual evidence and its role in problem understanding. Together with node identification, this step completes the construction of a coherent, problem-centric image graph.

\begin{figure}[!t]
    \centering
    \includegraphics[width=0.95\linewidth]{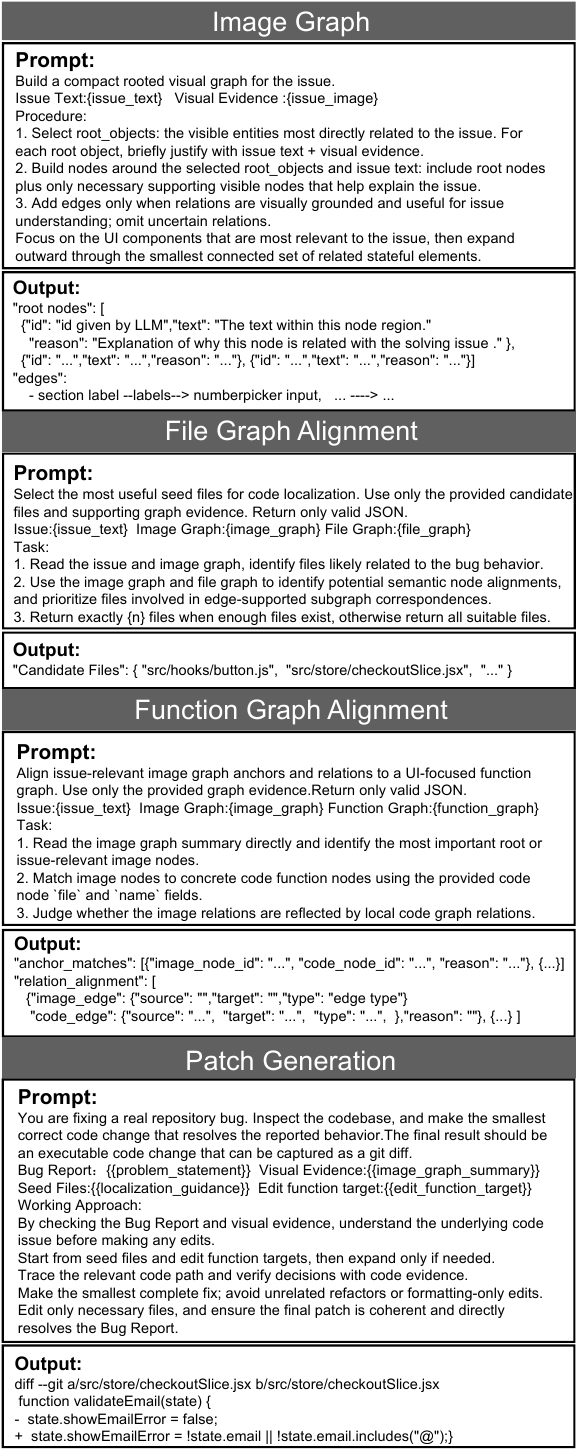}
    \caption{Stage-wise Simplified Prompt for GALA}
    \label{prompt}
\end{figure}

\subsection{File-level Alignment}

Given the constructed image graph, the goal of file-level alignment is to perform coarse cross-modal grounding at the repository level by identifying a compact set of files that are most likely related to the observed visual issue. As illustrated in Figure~\ref{method}, this stage progressively reduces the search space from the full repository to a dependency-consistent seed file set by combining repository structure, issue-relevant visual semantics, and lightweight inter-file relations. This coarse grounding step is designed to improve downstream localization efficiency while preserving the files most likely to contain the bug-related implementation.

\textbf{Repository Grounding.}
We construct a repository snapshot for each target commit, and recursively traverse the repository to collect file paths. During traversal, we filter out irrelevant directories (e.g., \textit{.git}, \textit{node\_modules}, \textit{dist}) and test related paths, and retain only repair relevant file types such as JavaScript/TypeScript files, style sheets, configuration files, and documentation. The retained file paths are then organized into a hierarchical directory structure (\textit{repo tree}), which serves as a structured prompt input for subsequent candidate and seed file selection, enabling the model to reason over repository organization rather than treating files as an unstructured collection.

\textbf{Visual Semantic Candidate Retrieval.}
However, repository structure alone does not indicate which files are relevant to the visual issue. To introduce problem-specific signals, we perform a semantics-guided candidate retrieval step by conditioning the model on the problem statement, the hierarchical \textit{repo tree} derived from the filtered repository snapshot, and compact structured summaries of all image graphs in the instance. These summaries retain key nodes, edges, and issue-relevant reasons from the original image graphs, while serializing them into a compact text-friendly form that can be readily consumed by the text LLM, reducing the overhead of passing full JSON graphs as well as the context length and inference cost. The problem statement provides behavioral descriptions, and the repo tree constrains retrieval within the filtered repository structure. Based on these complementary inputs, the model retrieves a fixed-size \textit{candidate file set} that is semantically consistent with the observed issue. This stage focuses on semantically grounding the search space before refinement, but does not yet explicitly dependencies between files; such structural evidence is introduced in the subsequent graph-based refinement step.

\textbf{Structure aware Refinement via File Graph Alignment.}
Within the same model call, we further introduce structural constraints by constructing a \textit{candidate file graph} over the retrieved candidate file list, where nodes correspond to candidate files and edges represent inter-file dependencies derived from static import relationships. This graph enables structure aware reasoning beyond purely semantic matching. Building on this representation, the model refines the retrieved candidate file list into a compact set of \textit{seed files} through a graph guided alignment step. Specifically, it jointly considers the problem statement, the image graph, and the candidate file graph, and prioritizes candidate files that are not only semantically relevant to the issue but also supported by consistent structural evidence. In particular, candidate files connected by supporting dependency relations in the file graph are favored over isolated matches. The model then returns a small fixed size set of seed files, which forms a dependency consistent subgraph likely to contain the bug and serves as the input for subsequent function level alignment.

\subsection{Function-level Alignment}
While file level alignment narrows the search space to a small set of seed files, it remains insufficient for precise localization, as individual files often contain multiple executable units with different responsibilities. We define alignment as identifying semantically compatible and structurally supported correspondences between issue-relevant image nodes and candidate code nodes within the induced code graph. To further bridge the gap between visual symptoms and executable code, we perform function-level alignment to align image graphs with the code structure and identify concrete edit targets. Formally, we define the function graph as \(G_c = (V_c, E_c)\), where nodes \(V_c\) correspond to functions, component-level units, or class methods, and edges \(E_c\) encode UI-relevant interactions such as rendering, data propagation, and state updates.

\textbf{UI-oriented Function Graph Construction.}
Given the selected seed files, we construct a closed-domain UI-oriented function graph within the seed files. We parse JavaScript/TypeScript source files in the seed set and extract three types of callable program units using fixed syntactic patterns, including function declarations, variable-assigned functions / arrow functions, and class methods, and normalize them into function-level nodes with identifiers, types, names, and file-level provenance. We then construct directed edges using regex/heuristic patterns with fixed triggers over function bodies, covering UI-relevant relations such as \textit{renders}, \textit{calls}, \textit{reads\_state}, \textit{writes\_state}, \textit{passes\_props}, and \textit{applies\_style}. For each source node, candidate target names are extracted from JSX component tags, call expressions, and state- or style-related symbols, and resolved in a fixed order: exact-name match within the same file, short-name match within the same file, and finally lookup through a seed-scoped symbol index. Edges are added when the referenced targets can be resolved to nodes inside the same seed-scoped graph; otherwise they are discarded, enforcing the closed-domain constraint. The resulting graph \(G_c = (V_c, E_c)\) provides a compact structural view of rendering logic, component interaction, and state transitions, and serves as the code-side input for subsequent cross-modal alignment and edit target identification.

\textbf{Cross-modal Graph Alignment.}
Given the constructed function graph, we perform function-level alignment with the image graph through a structured cross-modal reasoning step. The model is conditioned on the problem statement, the image graph summary, and the function graph, as specified in the simplified prompt shown in Fig.~\ref{prompt}, so that node semantics and typed relations are preserved for joint reasoning over visual and code structures. Pure semantic similarity is often insufficient for reliable grounding, as visually similar elements may correspond to structurally unrelated code components. We therefore restrict alignment to issue-relevant image nodes and candidate function nodes in the seed-induced graph, and retain a correspondence only when semantic compatibility is supported by relation-consistent neighborhood evidence. Concretely, the model jointly grounds image nodes to semantically related function nodes and verifies whether image-side relations are supported by compatible UI-oriented interactions in the function graph. The resulting alignment output is a small aligned subgraph that includes matched nodes, relation-supported correspondences, and concise rationales for downstream target selection. By combining entity grounding with relation-aware verification, this design filters out structurally inconsistent matches and yields more reliable localization evidence than retrieval based solely on independent semantic similarity.

\textbf{Edit Target Identification.}
Based on the alignment results, the model further identifies a small set of edit targets from the aligned subgraph within the same reasoning process to guide patch generation. Not all aligned functions are equally responsible for the observed issue, and directly modifying all candidates would introduce unnecessary changes and reduce reliability. Therefore, instead of relying on similarity alone, the model prioritizes functions that are strongly supported by both node-level and relation-level alignment evidence, and organizes them into primary, secondary, and contextual roles according to their expected repair relevance. The resulting target set provides a structured and interpretable bridge from multimodal alignment signals to actionable code edits, enabling focused and efficient bug fixing.

\subsection{Graph-guided Patch Generation}

After hierarchical localization, the system shifts from identifying issue-relevant code regions to generating an executable fix. We formulate this stage as \textit{graph-guided patch generation}, where repair is initialized from the aligned results of the previous stages rather than from unconstrained repository exploration. Specifically, the repair agent is provided with the image graph summary, the seed files from file-level alignment, and the edit targets from function-level alignment. The image graph summary preserves the visual or behavioral symptom to be resolved, while the seed files and edit targets define where inspection and modification should begin. The agent expands beyond these aligned regions only when local dependency evidence indicates that adjacent components must also be considered.

\textbf{Localized Repair Space.}
The outputs of file-level and function-level alignment jointly define a localized repair space. Seed files specify a compact set of issue-relevant files for prioritized inspection, while edit targets identify the classes, functions, or modules most directly implicated by the alignment results. Together, they reduce the ambiguity of repository-wide search and preserve the most relevant code context.

\textbf{Localized Dependency-aware Reasoning and Target Prioritization.}
Although the aligned repair space narrows the search scope, a bug may still involve interactions among nearby components. The agent therefore inspects short, issue-relevant dependency paths around the seed files and edit targets, considering relations such as function calls, component rendering, state access, and data propagation. This helps determine whether the anomaly is introduced locally or through a closely related upstream or downstream component. In addition, the agent uses the target roles produced during alignment (e.g., primary, secondary, and contextual) to prioritize inspection and editing: primary targets are examined first, secondary targets are considered when coordinated changes are needed, and contextual targets mainly support reasoning unless direct modification becomes necessary.

\textbf{Patch Generation.}
Within the same repair process, the agent generates a candidate patch from the localized repair space, beginning with the primary edit targets whenever possible. It is encouraged to preserve surrounding logic and avoid broad refactoring unless dependency evidence indicates that coordinated modifications are necessary. After code modification, the repository state for each instance is exported as a standardized patch file. Rather than producing an arbitrary textual patch, the modified files are first staged with \texttt{git add -A}, and the final patch is then exported using \texttt{git diff --cached} to ensure a valid and reproducible diff format. The resulting patch file is subsequently submitted to the SWE-Bench Multimodal evaluation pipeline, where final correctness is determined by the downstream execution and test protocol.

\section{Experiments}

\begin{table*}[t]
\centering
\renewcommand{\arraystretch}{1.1}
\caption{Comparison of Pass@1 resolve rate (\%) on SWE-Bench Multimodal.}
\resizebox{0.6\textwidth}{!}{
\begin{tabular}{lccc}
\toprule
\textbf{Method} & \textbf{Base model} & \textbf{\%Resolved} & \textbf{\#Resolved} \\
\midrule
SWE-Agent Multimodal & GPT-4o & 12.19 & 63 \\
Agentless Lite & Claude-3.5 Sonnet & 25.34 & 131 \\
Zencoder & Claude-3.5 Sonnet & 30.56 & 158 \\
OpenHands-Versa & Claude-Sonnet 4 & 34.43 & 178 \\
GUIRepair & Qwen3.5-122B-A10B & 34.82 & 180 \\
SVRepair & Qwen3.5-35B-A3B & 32.10 & 166 \\
SVRepair & Qwen3.5-122B-A10B & 33.66 & 174 \\
\midrule
GALA & Qwen3.5-35B-A3B & 33.66 & 174 \\
GALA & Qwen3.5-122B-A10B & \textbf{35.40} & \textbf{183} \\
\bottomrule
\end{tabular}
}
\label{tab:main_result}
\end{table*}

In this section, we conduct comprehensive experiments to evaluate the effectiveness of GALA on the SWE-Bench Multimodal benchmark. We begin by describing the experimental setup, including the benchmark, evaluation protocol, and baseline methods for comparison. We then report the main results in terms of end-to-end repair performance, followed by a detailed analysis of localization accuracy at both file-level and function-level. To further validate the robustness of our approach, we evaluate GALA under different model scales (Qwen3.5-35B-A3B and Qwen3.5-122B-A10B), examining how structured cross-modal alignment affects performance across varying model capacities. Finally, we conduct ablation studies to analyze the contribution of each component and to investigate the impact of code graph granularity. Together, these experiments provide a comprehensive assessment of both the effectiveness and the underlying mechanisms of our method.

\subsection{Implementation Details}
We implement GALA using large language models from the Qwen3.5 family, including \textit{Qwen3.5-35B-A3B} and \textit{Qwen3.5-122B-A10B}. These models are used throughout the full pipeline, including visual understanding, code localization, and patch generation. Specifically, the model serves as a vision-language model (VLM) for image graph construction and as a text model for subsequent reasoning and code-related tasks. This design helps maintain consistent semantic representations across stages. During file-level localization, we retrieve up to 10 candidate files and further select 5 seed files for downstream alignment. All generation and reasoning processes use a temperature of 0.0 to ensure deterministic outputs. For patch generation, we employ a cfuse-based agent guided by the structured alignment outputs. For fair baseline comparison, we re-implement GUIRepair and SVRepair using our deployed models while strictly following the implementation details reported in their original papers. The 35B model is served on 2 GPUs, while the 122B model is deployed on 8 GPUs. We set the maximum number of workers to 12 to parallelize inference. All experiments are conducted on a multi-GPU server equipped with NVIDIA A800-SXM4-80GB GPUs.

\subsection{Experiment Setup}

\textbf{\emph{Benchmark Selection}}\newline
We evaluate GALA on \textbf{SWE-Bench Multimodal} (SWE-Bench M) \cite{yangswe}, a benchmark designed to assess the ability of AI agents to resolve real-world software issues requiring both code reasoning and visual understanding. It extends the original SWE-Bench by introducing 619 tasks from 17 widely-used JavaScript repositories focused on user-facing and visually intensive applications, such as web interfaces, data visualization, and interactive systems. Each task includes both natural language descriptions and visual inputs (e.g., screenshots, diagrams, or rendered outputs), which are often essential for identifying the underlying problem. Compared to SWE-Bench, which is limited to Python repositories and primarily text-based inputs, SWE-Bench M introduces greater complexity through multimodal reasoning and cross-language code modifications. The benchmark is divided into a test set of 517 instances and a development set of 102 instances. We report results on the test set for overall evaluation and use the development set for model selection and analysis.

\noindent\textbf{\emph{Baseline Selection}}\newline
We compare GALA with several recent state-of-the-art approaches that have demonstrated strong performance on the SWE-Bench Multimodal benchmark. In particular, we include the top-performing methods reported in prior work as primary baselines. For the two strongest baselines, we further conduct additional experiments using our deployed model, \textit{Qwen3.5-122B-A10B}, in addition to their originally reported models. This allows for a more controlled and fair comparison by evaluating these methods under a consistent model setting, while also preserving their original reported performance for reference.

\subsection{Evaluation Results on SWE-Bench Multimodal}

\textbf{Main Experiment.}
We evaluate the overall effectiveness of GALA on the SWE-Bench Multimodal benchmark, with results shown in Table~\ref{tab:main_result}. GALA achieves the best performance among all compared methods, reaching a resolution rate of 35.40\%, outperforming strong multimodal baselines such as GUIRepair (34.82\%) and OpenHands-Versa (34.43\%). Notably, under the same base model setting (Qwen3.5-122B-A10B), GALA surpasses SVRepair (33.66\%) by a clear margin, demonstrating that the improvement stems from our proposed image graph–code graph alignment rather than model scaling. Furthermore, GALA significantly outperforms earlier approaches such as Agentless Lite (25.34\%) and SWE-Agent Multimodal (12.19\%), highlighting the importance of structured cross-modal reasoning in multimodal program repair. We also observe that recent multimodal approaches, including GUIRepair and SVRepair, exhibit relatively modest gains over prior methods, suggesting that performance improvements in this benchmark are inherently incremental and further underscoring the effectiveness of our structured alignment strategy. Overall, these results validate that GALA provides a more accurate and structurally grounded localization and repair paradigm, achieving state-of-the-art performance on SWE-Bench Multimodal.

\textbf{Localization Performance.}
We further evaluate the localization capability of GALA at both file-level and function-level on the validation split of SWE-Bench Multimodal, since the test split does not expose gold patches and thus does not support localization evaluation. As shown in Table~\ref{tab:loc}, GALA consistently achieves the best performance across all settings. Under the 122B model, GALA reaches 29.22\% file-level recall and 17.14\% function-level recall, outperforming SVRepair (28.71\% / 16.25\%) and GUIRepair (24.3\% / 12.4\%). More importantly, the improvement is even more pronounced under the smaller 35B model: GALA achieves 28.22\% file-level and 15.88\% function-level recall, exceeding SVRepair (25.50\% / 13.63\%) by a larger margin compared to the 122B setting. This observation suggests that our structured graph alignment provides stronger guidance when model capacity is limited, enabling more effective localization even with smaller models. These results further confirm that GALA improves not only final repair success but also the underlying localization quality, which is critical for multimodal bug fixing.

\begin{table}[t]
\centering
\renewcommand{\arraystretch}{1.1}
\caption{File-level and function-level localization recall (\%) on SWE-Bench M.}
\resizebox{0.9\columnwidth}{!}{
\begin{tabular}{lcc}
\toprule
\textbf{Method} & \textbf{File-level} & \textbf{Function-level} \\
\midrule
GUIRepair(35b)  &  21.78  &  10.17  \\
GUIRepair(122b)  &  24.3  &  12.4  \\
SVRepair(35b)   &  25.50  &  13.63  \\
SVRepair(122b)   &  28.71  &  16.25  \\
GALA(35b)       &  28.22  &  15.88  \\
GALA(122b)       &  \textbf{29.22}  &  \textbf{17.14}  \\
\bottomrule
\end{tabular}
}
\label{tab:loc}
\end{table}

\subsection{Ablation Studies}

\begin{table}[t]
\centering
\renewcommand{\arraystretch}{1.1}
\caption{Ablation study on different components of our method.}
\resizebox{1\linewidth}{!}{
\begin{tabular}{ccc|c}
\toprule
\textbf{Image Graph} & \textbf{Code Graph} & \textbf{Alignment} & \textbf{Resolved (\%)} \\
\midrule
-- & -- & -- & 33.66 \\
\checkmark & -- & -- & 34.43 \\
\checkmark & \checkmark & -- & 34.43 \\
\midrule
\checkmark & \checkmark & \checkmark & \textbf{35.40} \\
\bottomrule
\end{tabular}
}
\label{tab:ablation1}
\end{table}

\begin{table}[t]
\centering
\renewcommand{\arraystretch}{1.1}
\caption{Ablation study on different levels of code graph granularity.}
\resizebox{1\linewidth}{!}{
\begin{tabular}{cc|c}
\toprule
\textbf{File-level Graph} & \textbf{Function-level Graph} & \textbf{Resolved (\%)} \\
\midrule
-- & -- & 34.43 \\
\checkmark & -- & 34.62 \\
\checkmark & \checkmark & \textbf{35.40} \\
\bottomrule
\end{tabular}
}
\label{tab:ablation2}
\end{table}

\textbf{Ablation on core components.}  
To evaluate the contribution of each component in our framework, we conduct a systematic ablation study on SWE-Bench Multimodal, as shown in Table~\ref{tab:ablation1}. We consider four incremental configurations, implemented by selectively enabling or disabling structured components while keeping all other settings unchanged. We start from a text-only setting without structured visual or code representations, achieving 33.66\%. In this setting, the model does not utilize image inputs and skips the candidate file retrieval stage, instead directly selecting seed files based on the problem description and repository context, with the model guided to focus on these seed files during reasoning. Introducing the image graph improves performance to 34.43\%, indicating that structured visual information provides useful signals beyond pure text-based reasoning. Adding the code graph alone on top of the image graph does not yield further gains (34.43\%), suggesting that code-side structure without explicit cross-modal grounding is insufficient to improve localization. In contrast, enabling cross-modal alignment between the image graph and code graph leads to a substantial improvement, reaching 35.40\%. This result highlights that the key performance gain comes from explicitly aligning visual and code structures, rather than modeling them independently.

\textbf{Ablation on code graph granularity.}  
We further investigate the impact of code graph granularity, with results summarized in Table~\ref{tab:ablation2}. We consider configurations with different levels of structural granularity by selectively enabling file-level and finer-grained code graph representations while keeping all other components unchanged. Starting from a setting with image graph and file-level reasoning only (34.43\%), introducing a file-level code graph improves performance to 34.62\%, demonstrating that coarse-grained structural information is effective for narrowing down candidate regions. Further incorporating a finer-grained function-level graph leads to a higher performance of 35.40\%, indicating that fine-grained representations provide more precise localization signals. This progression shows that multi-level structural modeling provides complementary benefits, where file-level graphs offer global structural context while finer-grained representations enable more accurate reasoning over local code regions.

\section{Conclusion}

In this paper, we presented \textbf{GALA}, a multimodal automated program repair framework that formulates bug localization as a hierarchical cross-modal graph alignment problem. By modeling visual structures with an image graph and code dependencies with multi-level code graphs, GALA enables structured reasoning between visual observations and executable code. Through file- and function-level alignment, our approach enforces semantic and relational consistency, yielding more accurate and interpretable localization. Experiments on SWE-bench Multimodal show that GALA outperforms existing methods, validating the effectiveness of structure-aware alignment in multimodal APR. Consistent gains across model scales further indicate that our framework provides robust guidance beyond specific model capacities. This work highlights the importance of moving beyond implicit textual representations toward explicit structural reasoning, opening new directions for graph-based multimodal software engineering.

\section{Limitations}

Despite its effectiveness, our method has several limitations. Due to limited context capacity of current LLMs, fine-grained alignment between visual elements and line-level code remains challenging, and we instead adopt hierarchical alignment from files to functions. Consequently, the method depends on repository organization and may degrade under ambiguous naming or modular structures, though partially mitigated by increasingly standardized, AI-assisted development. In addition, our framework introduces structured intermediate representations as external reasoning scaffolds to simplify multimodal understanding. As LLMs and agent systems advance, such structures may become internalized, enabling more direct end-to-end reasoning. Balancing explicit structural guidance with increasingly powerful implicit reasoning remains an important direction for future work.


\bibliographystyle{ACM-Reference-Format}
\bibliography{sample-base}

@article{samir2026improved,
  title={Improved Bug Localization with AI Agents Leveraging Hypothesis and Dynamic Cognition},
  author={Samir, Asif Mohammed and Rahman, Mohammad Masudur},
  journal={arXiv preprint arXiv:2601.12522},
  year={2026}
}

@inproceedings{jiang2025issue,
  title={Issue Localization via LLM-Driven Iterative Code Graph Searching},
  author={Jiang, Zhonghao and Ren, Xiaoxue and Yan, Meng and Jiang, Wei and Li, Yong and Liu, Zhongxin},
  booktitle={2025 40th IEEE/ACM International Conference on Automated Software Engineering (ASE)},
  pages={3034--3045},
  year={2025},
  organization={IEEE}
}

@article{xu2026learning,
  title={Learning Adaptive Parallel Execution for Efficient Code Localization},
  author={Xu, Ke and Xiao, Siyang and Liang, Ming and Yu, Yichen and Wang, Zhixiang and Xu, Jingxuan and Chen, Dajun and Jiang, Wei and Li, Yong},
  journal={arXiv preprint arXiv:2601.19568},
  year={2026}
}

@article{liu2025graphlocator,
  title={GraphLocator: Graph-guided Causal Reasoning for Issue Localization},
  author={Liu, Wei and Peng, Chao and Gao, Pengfei and Liu, Aofan and Zhang, Wei and Zhao, Haiyan and Jin, Zhi},
  journal={arXiv preprint arXiv:2512.22469},
  year={2025}
}

@inproceedings{chen2025locagent,
  title={Locagent: Graph-guided llm agents for code localization},
  author={Chen, Zhaoling and Tang, Robert and Deng, Gangda and Wu, Fang and Wu, Jialong and Jiang, Zhiwei and Prasanna, Viktor and Cohan, Arman and Wang, Xingyao},
  booktitle={Proceedings of the 63rd Annual Meeting of the Association for Computational Linguistics (Volume 1: Long Papers)},
  pages={8697--8727},
  year={2025}
}

@inproceedings{batole2025llm,
  title={An LLM-Based Agent-Oriented Approach for Automated Code Design Issue Localization},
  author={Batole, Fraol and OBrien, David and Nguyen, Tien and Dyer, Robert and Rajan, Hridesh},
  booktitle={2025 IEEE/ACM 47th International Conference on Software Engineering (ICSE)},
  pages={637--637},
  year={2025},
  organization={IEEE Computer Society}
}

@article{li2025swe,
  title={Swe-debate: Competitive multi-agent debate for software issue resolution},
  author={Li, Han and Shi, Yuling and Lin, Shaoxin and Gu, Xiaodong and Lian, Heng and Wang, Xin and Jia, Yantao and Huang, Tao and Wang, Qianxiang},
  journal={arXiv preprint arXiv:2507.23348},
  year={2025}
}

@article{wang2025extracting,
  title={Extracting Conceptual Knowledge to Locate Software Issues},
  author={Wang, Ying and Mao, Wenjun and Wang, Chong and Zhou, Zhenhao and Zhou, Yicheng and Zhao, Wenyun and Lou, Yiling and Peng, Xin},
  journal={arXiv preprint arXiv:2509.21427},
  year={2025}
}

@article{wang2025improving,
  title={Improving code localization with repository memory},
  author={Wang, Boshi and Xu, Weijian and Li, Yunsheng and Gao, Mei and Xie, Yujia and Sun, Huan and Chen, Dongdong},
  journal={arXiv preprint arXiv:2510.01003},
  year={2025}
}

@article{reddy2025swerank,
  title={Swerank: Software issue localization with code ranking},
  author={Reddy, Revanth Gangi and Suresh, Tarun and Doo, JaeHyeok and Liu, Ye and Nguyen, Xuan Phi and Zhou, Yingbo and Yavuz, Semih and Xiong, Caiming and Ji, Heng and Joty, Shafiq},
  journal={arXiv preprint arXiv:2505.07849},
  year={2025}
}

@inproceedings{soni2025coding,
  title={Coding Agents with Multimodal Browsing are Generalist Problem Solvers},
  author={Soni, Aditya Bharat and Li, Boxuan and Wang, Xingyao and Chen, Valerie and Neubig, Graham},
  booktitle={ICML 2025 Workshop on Computer Use Agents},
  year={2025}
}

@article{huang2025seeing,
  title={Seeing is fixing: Cross-modal reasoning with multimodal llms for visual software issue fixing},
  author={Huang, Kai and Zhang, Jian and Xie, Xiaofei and Chen, Chunyang},
  journal={arXiv preprint arXiv:2506.16136},
  year={2025}
}

@article{tang2026svrepair,
  title={SVRepair: Structured Visual Reasoning for Automated Program Repair},
  author={Tang, Xiaoxuan and Wang, Jincheng and Luo, Liwei and Xu, Jingxuan and Zhou, Sheng and Chen, Dajun and Jiang, Wei and Li, Yong},
  journal={arXiv preprint arXiv:2602.06090},
  year={2026}
}

@inproceedings{yangswe,
  title={SWE-bench Multimodal: Do AI Systems Generalize to Visual Software Domains?},
  author={Yang, John and Jimenez, Carlos E and Zhang, Alex L and Lieret, Kilian and Yang, Joyce and Wu, Xindi and Press, Ori and Muennighoff, Niklas and Synnaeve, Gabriel and Narasimhan, Karthik R and others},
  booktitle={The Thirteenth International Conference on Learning Representations},
  year={2025}
}

@inproceedings{jiang2023impact,
  title={Impact of code language models on automated program repair},
  author={Jiang, Nan and Liu, Kevin and Lutellier, Thibaud and Tan, Lin},
  booktitle={2023 IEEE/ACM 45th International Conference on Software Engineering (ICSE)},
  pages={1430--1442},
  year={2023},
  organization={IEEE}
}

@inproceedings{xia2023automated,
  title={Automated program repair in the era of large pre-trained language models},
  author={Xia, Chunqiu Steven and Wei, Yuxiang and Zhang, Lingming},
  booktitle={2023 IEEE/ACM 45th International Conference on Software Engineering (ICSE)},
  pages={1482--1494},
  year={2023},
  organization={IEEE}
}

@inproceedings{wu2023effective,
  title={How effective are neural networks for fixing security vulnerabilities},
  author={Wu, Yi and Jiang, Nan and Pham, Hung Viet and Lutellier, Thibaud and Davis, Jordan and Tan, Lin and Babkin, Petr and Shah, Sameena},
  booktitle={Proceedings of the 32nd ACM SIGSOFT International Symposium on Software Testing and Analysis},
  pages={1282--1294},
  year={2023}
}

@inproceedings{xia2022less,
  title={Less training, more repairing please: revisiting automated program repair via zero-shot learning},
  author={Xia, Chunqiu Steven and Zhang, Lingming},
  booktitle={Proceedings of the 30th ACM Joint European Software Engineering Conference and Symposium on the Foundations of Software Engineering},
  pages={959--971},
  year={2022}
}

@inproceedings{fan2023automated,
  title={Automated repair of programs from large language models},
  author={Fan, Zhiyu and Gao, Xiang and Mirchev, Martin and Roychoudhury, Abhik and Tan, Shin Hwei},
  booktitle={2023 IEEE/ACM 45th International Conference on Software Engineering (ICSE)},
  pages={1469--1481},
  year={2023},
  organization={IEEE}
}

@inproceedings{zhao2024enhancing,
  title={Enhancing automated program repair with solution design},
  author={Zhao, Jiuang and Yang, Donghao and Zhang, Li and Lian, Xiaoli and Yang, Zitian and Liu, Fang},
  booktitle={Proceedings of the 39th IEEE/ACM International Conference on Automated Software Engineering},
  pages={1706--1718},
  year={2024}
}

@article{yang2024swe,
  title={Swe-agent: Agent-computer interfaces enable automated software engineering},
  author={Yang, John and Jimenez, Carlos E and Wettig, Alexander and Lieret, Kilian and Yao, Shunyu and Narasimhan, Karthik and Press, Ofir},
  journal={Advances in Neural Information Processing Systems},
  volume={37},
  pages={50528--50652},
  year={2024}
}

@article{xia2024agentless,
  title={Agentless: Demystifying llm-based software engineering agents},
  author={Xia, Chunqiu Steven and Deng, Yinlin and Dunn, Soren and Zhang, Lingming},
  journal={arXiv preprint arXiv:2407.01489},
  year={2024}
}

@inproceedings{zhang2024autocoderover,
  title={Autocoderover: Autonomous program improvement},
  author={Zhang, Yuntong and Ruan, Haifeng and Fan, Zhiyu and Roychoudhury, Abhik},
  booktitle={Proceedings of the 33rd ACM SIGSOFT International Symposium on Software Testing and Analysis},
  pages={1592--1604},
  year={2024}
}

@inproceedings{ma2025alibaba,
  title={Alibaba lingmaagent: Improving automated issue resolution via comprehensive repository exploration},
  author={Ma, Yingwei and Yang, Qingping and Cao, Rongyu and Li, Binhua and Huang, Fei and Li, Yongbin},
  booktitle={Proceedings of the 33rd ACM International Conference on the Foundations of Software Engineering},
  pages={238--249},
  year={2025}
}

@article{antoniades2024swe,
  title={Swe-search: Enhancing software agents with monte carlo tree search and iterative refinement},
  author={Antoniades, Antonis and {\"O}rwall, Albert and Zhang, Kexun and Xie, Yuxi and Goyal, Anirudh and Wang, William},
  journal={arXiv preprint arXiv:2410.20285},
  year={2024}
}

@article{xia2025demystifying,
  title={Demystifying llm-based software engineering agents},
  author={Xia, Chunqiu Steven and Deng, Yinlin and Dunn, Soren and Zhang, Lingming},
  journal={Proceedings of the ACM on Software Engineering},
  volume={2},
  number={FSE},
  pages={801--824},
  year={2025},
  publisher={ACM New York, NY, USA}
}

@article{jimenez2023swe,
  title={Swe-bench: Can language models resolve real-world github issues?},
  author={Jimenez, Carlos E and Yang, John and Wettig, Alexander and Yao, Shunyu and Pei, Kexin and Press, Ofir and Narasimhan, Karthik},
  journal={arXiv preprint arXiv:2310.06770},
  year={2023}
}

@inproceedings{lam2017bug,
  title={Bug localization with combination of deep learning and information retrieval},
  author={Lam, An Ngoc and Nguyen, Anh Tuan and Nguyen, Hoan Anh and Nguyen, Tien N},
  booktitle={2017 IEEE/ACM 25th International Conference on Program Comprehension (ICPC)},
  pages={218--229},
  year={2017},
  organization={IEEE}
}

@inproceedings{ciborowska2022fast,
  title={Fast changeset-based bug localization with BERT},
  author={Ciborowska, Agnieszka and Damevski, Kostadin},
  booktitle={Proceedings of the 44th international conference on software engineering},
  pages={946--957},
  year={2022}
}

@article{chakraborty2025blaze,
  title={BLAZE: Cross-language and cross-project bug localization via dynamic chunking and hard example learning},
  author={Chakraborty, Partha and Alfadel, Mahmoud and Nagappan, Meiyappan},
  journal={IEEE Transactions on Software Engineering},
  year={2025},
  publisher={IEEE}
}

@article{ali2023automated,
  title={Automated software bug localization enabled by meta-heuristic-based convolutional neural network and improved deep neural network},
  author={Ali, Waqas and Bo, Lili and Sun, Xiaobing and Wu, Xiaoxue and Memon, Saifullah and Siraj, Saima and Ashton, Ann Suwaree},
  journal={Expert Systems with Applications},
  volume={232},
  pages={120562},
  year={2023},
  publisher={Elsevier}
}

@inproceedings{huo2017enhancing,
  title={Enhancing the Unified Features to Locate Buggy Files by Exploiting the Sequential Nature of Source Code.},
  author={Huo, Xuan and Li, Ming},
  booktitle={IJCAI},
  pages={1909--1915},
  year={2017}
}

@inproceedings{wang2023rap,
  title={Rap-gen: Retrieval-augmented patch generation with codet5 for automatic program repair},
  author={Wang, Weishi and Wang, Yue and Joty, Shafiq and Hoi, Steven CH},
  booktitle={Proceedings of the 31st ACM Joint European Software Engineering Conference and Symposium on the Foundations of Software Engineering},
  pages={146--158},
  year={2023}
}

@inproceedings{huang2025template,
  title={Template-Guided Program Repair in the Era of Large Language Models.},
  author={Huang, Kai and Zhang, Jian and Meng, Xiangxin and Liu, Yang},
  booktitle={ICSE},
  pages={1895--1907},
  year={2025}
}

@inproceedings{xia2023plastic,
  title={The plastic surgery hypothesis in the era of large language models},
  author={Xia, Chunqiu Steven and Ding, Yifeng and Zhang, Lingming},
  booktitle={2023 38th IEEE/ACM International Conference on Automated Software Engineering (ASE)},
  pages={522--534},
  year={2023},
  organization={IEEE}
}

@inproceedings{bouzenia2025repairagent,
  title={Repairagent: An autonomous, llm-based agent for program repair},
  author={Bouzenia, Islem and Devanbu, Premkumar and Pradel, Michael},
  booktitle={2025 IEEE/ACM 47th International Conference on Software Engineering (ICSE)},
  pages={2188--2200},
  year={2025},
  organization={IEEE}
}

@inproceedings{lee2025unidebugger,
  title={Unidebugger: Hierarchical multi-agent framework for unified software debugging},
  author={Lee, Cheryl and Xia, Chunqiu Steven and Yang, Longji and Huang, Jen-tse and Zhu, Zhouruixing and Zhang, Lingming and Lyu, Michael R},
  booktitle={Proceedings of the 2025 Conference on Empirical Methods in Natural Language Processing},
  pages={18248--18277},
  year={2025}
}

@inproceedings{yin2024thinkrepair,
  title={Thinkrepair: Self-directed automated program repair},
  author={Yin, Xin and Ni, Chao and Wang, Shaohua and Li, Zhenhao and Zeng, Limin and Yang, Xiaohu},
  booktitle={Proceedings of the 33rd ACM SIGSOFT International Symposium on Software Testing and Analysis},
  pages={1274--1286},
  year={2024}
}

@inproceedings{zhang2023gamma,
  title={Gamma: Revisiting template-based automated program repair via mask prediction},
  author={Zhang, Quanjun and Fang, Chunrong and Zhang, Tongke and Yu, Bowen and Sun, Weisong and Chen, Zhenyu},
  booktitle={2023 38th IEEE/ACM International Conference on Automated Software Engineering (ASE)},
  pages={535--547},
  year={2023},
  organization={IEEE}
}

@inproceedings{yang2024cref,
  title={Cref: An llm-based conversational software repair framework for programming tutors},
  author={Yang, Boyang and Tian, Haoye and Pian, Weiguo and Yu, Haoran and Wang, Haitao and Klein, Jacques and Bissyand{\'e}, Tegawend{\'e} F and Jin, Shunfu},
  booktitle={Proceedings of the 33rd ACM SIGSOFT International Symposium on Software Testing and Analysis},
  pages={882--894},
  year={2024}
}

@inproceedings{peng2024domain,
  title={Domain knowledge matters: Improving prompts with fix templates for repairing python type errors},
  author={Peng, Yun and Gao, Shuzheng and Gao, Cuiyun and Huo, Yintong and Lyu, Michael},
  booktitle={Proceedings of the 46th ieee/acm international conference on software engineering},
  pages={1--13},
  year={2024}
}

@article{le2019automated,
  title={Automated program repair},
  author={Le Goues, Claire and Pradel, Michael and Roychoudhury, Abhik},
  journal={Communications of the ACM},
  volume={62},
  number={12},
  pages={56--65},
  year={2019},
  publisher={ACM New York, NY, USA}
}

@article{zhang2023survey,
  title={A survey of learning-based automated program repair},
  author={Zhang, Quanjun and Fang, Chunrong and Ma, Yuxiang and Sun, Weisong and Chen, Zhenyu},
  journal={ACM Transactions on Software Engineering and Methodology},
  volume={33},
  number={2},
  pages={1--69},
  year={2023},
  publisher={ACM New York, NY}
}

@article{zhang2024systematic,
  title={A systematic literature review on large language models for automated program repair},
  author={Zhang, Quanjun and Fang, Chunrong and Xie, Yang and Ma, YuXiang and Sun, Weisong and Yang, Yun and Chen, Zhenyu},
  journal={ACM Transactions on Software Engineering and Methodology},
  year={2024},
  publisher={ACM New York, NY}
}

@inproceedings{li2024mmcode,
  title={Mmcode: Benchmarking multimodal large language models for code generation with visually rich programming problems},
  author={Li, Kaixin and Tian, Yuchen and Hu, Qisheng and Luo, Ziyang and Huang, Zhiyong and Ma, Jing},
  booktitle={Findings of the Association for Computational Linguistics: EMNLP 2024},
  pages={736--783},
  year={2024}
}

@article{wang2025code,
  title={Code-vision: evaluating multimodal LLMs logic understanding and code generation capabilities},
  author={Wang, Hanbin and Zhou, Xiaoxuan and Xu, Zhipeng and Cheng, Keyuan and Zuo, Yuxin and Tian, Kai and Song, Jingwei and Lu, Junting and Hu, Wenhui and Liu, Xueyang},
  journal={arXiv preprint arXiv:2502.11829},
  year={2025}
}

@inproceedings{tan2017codeflaws,
  title={Codeflaws: a programming competition benchmark for evaluating automated program repair tools},
  author={Tan, Shin Hwei and Yi, Jooyong and Mechtaev, Sergey and Roychoudhury, Abhik and others},
  booktitle={2017 IEEE/ACM 39th International Conference on Software Engineering Companion (ICSE-C)},
  pages={180--182},
  year={2017},
  organization={IEEE}
}

@inproceedings{ouyang2024benchmarking,
  title={Benchmarking automated program repair: An extensive study on both real-world and artificial bugs},
  author={Ouyang, Yicheng and Yang, Jun and Zhang, Lingming},
  booktitle={Proceedings of the 33rd ACM SIGSOFT International Symposium on Software Testing and Analysis},
  pages={440--452},
  year={2024}
}

@inproceedings{lin2017quixbugs,
  title={QuixBugs: A multi-lingual program repair benchmark set based on the Quixey Challenge},
  author={Lin, Derrick and Koppel, James and Chen, Angela and Solar-Lezama, Armando},
  booktitle={Proceedings Companion of the 2017 ACM SIGPLAN international conference on systems, programming, languages, and applications: software for humanity},
  pages={55--56},
  year={2017}
}

\end{document}